\documentclass[12pt,a4,showpacs,superscriptaddress]{article}
\usepackage{epsfig}

\title{Metal Cluster's Effect on the Optical Properties of Cesium Bromide Thin Films}
\author{Kuldeep Kumar$^a$, P.Arun$^a$\footnote{e-mail: 
arunp92@physics.du.ac.in,
(T) +91 011 29258401, (F) +91 011 27666220}, Chhaya Ravi Kant$^b$, Bala
Krishna Juluri$^c$\\ \\
$^a$ Material Science Research Lab,\\ S.G.T.B. Khalsa College,\\ 
University of Delhi, Delhi-110 007, India\\
$^b$ Department of Applied Sciences, Indira Gandhi Institute of
Technology,\\ Guru Gobind Singh Indraprastha University, Delhi 110 006, 
India\\
$^c$ Department of Engineering Science and Mechanics,\\
The Pennsylvania State University,\\
Pennsylvania 16802, USA\\
}

\begin{document}
\maketitle

\vfil \eject
\begin{abstract}
Cesium Bromide films grown of glass substrates by thermal evaporation showed
interesting optical properties. The UV-visible absorption spectra showed
peaks which showed red shift with time. Structural and morphological studies
suggested decrease in grain size with time which was unusual. Theoretical
simulation shows the optical behaviour to be due to surface plasmon
resonance resulting from Cesium clyindrical rods embedded in the films.
\end{abstract}
{\bf PAC No: 68.55.-a, 71.20.Ps, 78.67.Bf}

\section{Introduction}

Alkali Halides (AH) have been extensively studied in the early years to
understand the variation in energy band structure with changing lattice
constant \cite{fuji}. The simple cubic crystal structures, experimentally 
determinable phase transitions and variation in properties when these 
samples are subjected to presure made them popular samples \cite{marco}. 
Also, AH are 
highly photon-absorbing materials in X-ray and Ultraviolet region, giving rise to 
color-centers \cite{mitc}-\cite{sever}. Cesium 
halides hence became popular in detectors and optoelectronics devices
\cite{sir}.
Cesium halides also showed excellent quantum efficiency. Infact, thin CsBr 
films deposited on thin metal layers have useful application as electron 
source and as protective layer of photocathode \cite{brakin}. 
However, AH films suffered due to their
high reactivity with atmospheric water vapor \cite{saiki} making it
difficult to maintain. In this manuscript we focus on Cesium Bromide, since
of all the alkali halides, it exhibits a relatively higher stability when 
exposure in air \cite{crc}. 

Recent interest in Alkali Halides is due to the appearance of Surface
Plasmon Resonance (SPR) peaks in the UV-visible region of its absorption 
spectrum, when noble and other metal nano-particle are embedded in it
\cite{nie}. 
However, we
have noticed SPR peaks developing in cesium iodide \cite{kapil} and cesium 
bromide (present work) thin films even with aging. We believe this to be a 
direct proof of the idea put forward 
by Emel'yanov and others \cite{emel, sea}. They claimed that the 
defect (such as F-centers) in 
alkali halides when provided the right conditions would lead to metal cluster 
formations. We believe it is these alkali
clusters that form the metal nanoparticles responsible for SPR formation. In 
the 
present work we report formation of Cesium cluster in CsBr thin films 
due to ageing effect and their effect on optical properties. 

\section{Experimental}
Cesium Bromide thin films were deposited by thermal evaporation on 
microscopy glass substrates. The starting material (powder) was obtained from 
HiMedia (Mumbai) and was of 99.98\% purity. Films depositons were 
carried out at room temperature with vacuum better than 
${\rm 1.4 \times 10^{-5}}$ Torr. The thickness of the films were 
monitered with the help of a Digital Thickness Moniter (DTM-106). One
part of the films were kept in desiccators and taken out only for 
characterization while the second part were kept outside, exposed to open
air. This was done to compare the ageing effect of these films maintained in
different conditions.

The structural and optical characterization the samples were done with 
Philips PW 3020 diffractometer and Systronics double beam UV-visible 
spectrophotometer (2202). Surface morphology and elemental composition 
of films were examined with Field-Emission Scaning Electron Microscopy 
(FE-SEM) FEI-Quanta 200F and the Energy Dispersive Analysis of X-rays (EDX) 
system attached with it. 

\section{Results and Discussion}

\subsection{Optical Studies}
The absorption spectra of Cesium Bromide films showed thickness dependence
with films thicker than 550~nm showing intense absorbance in the 
visible region. We believe that these peaks are due to surface plasmon 
resonance (SPR). SPR peaks results from the interaction between metal 
conduction electron and the electric field component of incident 
electromagnetic radiation \cite{vollmer}. This interaction leads to collective 
oscillations of free electrons on the metal surface if the metal particle's 
dimensions are smaller than or comparable to the incident wavelength of light. 
These peaks in CsBr's absorbance spectra are a 
result of metal nano-clusters of cesium existing within the films. Studies 
on Cesium Iodide too had revealed similar peaks in their UV-visible 
absorption spectra in our earlier studies \cite{kapil}.

However, what is intrigging in Cesium Bromide is the systematic variation in 
film property with time. UV-visible spectra recorded immediately after film 
fabrication showed a dominant peak around 500~nm. Not only did this peak 
red-shift with time but also its absorption intensity was also found to
decrease. The UV-visible absorption spectra (Fig~1) of 
a 840~nm thick sample taken at various intervals after film fabrication
exhibits this trend. This ageing effect of the sample is believed to depend
on the ambient atmosphere in which the sample was maintained, hence as
explained above, one set of the samples were maintained in a dessicator while 
the 
second set were maintained in a normal miscroscope slide box. As expected, both 
samples that started with identical UV-visible spectra showed 
different rate of variation. Those kept in dessicator aged slower.
The shift in peak position shows a linear trend with time and the rate of 
change was found to be film thickness dependent (fig~2a). Thicker films 
showed higher rate of shift as compared to the thinner films (fig~2b). 
Even the decrease in absorption showed systematic trend of variation with
time (fig~2c).

Both the SPR peak position and its intensity are strongly dependent on
the metal cluster's size, it's and that of the surrounding media dielectric 
constant. These results, hence, indicate some 
systematic and continuous changes in the film that require investigation. 
Below we focus our attention on how metal clusters arise in our samples and the 
structural and morphological changes which would explain the 
variations observed.

\subsection{Morphological, Structural \& Compositional Studies}

As can be seen from fig~3, sharp edged grains can be found uniformally 
scattered on the film surface. The micrographs of this figure compares two 
samples, (a) sample maintained in dessicator and (b) that kept outside. The 
grain density and grain size were determined from these micrographs using 
the software ImageJ. Below each micrograph histograms depicting the
grain size distribution in the micrograph are given. As can be seen, samples 
kept in dessicator showed slightly larger grains in the sample maintained in 
the dessicator as compared to that maintained outside (${\rm 1.00 \mu m}$ as
compared to ${\rm 0.95 \mu m}$). The grain density of samples kept in air 
(${\rm 0.91 \mu m^{-2}}$) was found to be nearly twice that of the counterpart 
kept in the dessicator (${\rm 0.52 \mu m^{-2}}$). The full width at half 
maxima (FWHM) of the Guassian fit to the histograms also reflect a narrower 
distribution for the samples maintained in the dessicator. Viewed in
conjucture these results suggest grains of the film split giving rise to 
smaller grains. Also, this process of grain breaking is encouraged in
samples kept in air. Figure~4 gives a suggestive sequence of ``grain 
division". Fig~4a shows two 
grains in close proximity that are inter-connected by vesticles. These
structures increase in length as the grains move apart. As the grains moves 
apart, the vesticle like structure is retained by one of the grains (see
fig~4b). These vesticles finally break off and fall into the background
(circles marked in fig~4c highlight this) leaving behind a smooth spherical
grain. EDX (Table~1) on vesticle and grains show them to be made up of cesium 
and cesium bromide respectively. These results would indicate removal of 
bromide from the surfaces of
the grains. The process is accelerated in the samples kept in air. The 
sublimation leaves behind cesium metal layer on the surface of the grain, 
resulting in a ``insulator-metal" ``core-shell" structure. Figure~5 is a
sample TEM micrograph which clearly shows distinct regions of core and
shell. As expected there is a large variation in the grain size. Also, seen 
distributed among the spherical shells are ``rod" like structures we have 
explained above. 

Fig~6(a) shows the Selected Area Electron Diffraction (SAED) taken on one of
the core of the ``core-shell" grains. The spots indicate the crystalline 
nature of the ``core". The major spots of the SAED are arranged on thwo 
distinct rings corresponding to the (110) and (211) peaks of CsBr as indexed 
in ASTM Card No~73-0391. Similar analysis of the rod region (fig~6b) shows it 
to be crystalline with three distinct ring corresponding to the (200), (331) 
and (220) planes of Cesium as given in the ASTM Card No~18-0325. 

X-ray 
Diffraction of the samples shows two major peaks of CsBr. The existence of
both CsBr and Cs in our samples are confimred by the broad peaks at
${\rm 2\theta \approx 29^o}$ and ${\rm \approx 52^o}$ (fig~7). These
peak positions match those given in ASTM Card No~73-0391 and 18-0325.
The grain sizes (Table 2) of CsBr core were also calculated from the X-ray 
diffraction (fig~7) peak's Full Width at Half Maxima (FWHM) using Scherrer 
formula \cite{cullity}. The smaller grain sizes indicate that the grain 
boundaries of the core, as viewed in the micrographs, enclose crystalline 
region along with amorphous CsBr. While the amount of free Cesium in the 
samples can be thought to be low, X-Ray diffraction pattern does show a
broad peak at ${\rm \approx 52^o}$, formed by merging of the (211) and (220)
peaks of CsBr and Cs. On deconvoluting these peaks, the average grain size of
free Cesium were also calculated. Grain size of CsBr and Cs were found to
decrease with increasing time. This trend is in agreement with those from
morphological studies. However, more importantly, the CsBr peaks were found
to have shifted to the right as compared to the peak positions given in the 
ASTM Card. This would indicate that the CsBr lattices are in a state of stress
with compressive forces acting on it. The stress in the film were calculated
after evaluating the strain using the relation \cite{pat}
\begin{eqnarray}
{\Delta d \over d} = {d_{obs}-d_{ASTM} \over d_{obs}}\label{strain}
\end{eqnarray}
where ${\rm d_{obs}}$ is the experimentally observed d-spacing and ${\rm
d_{ASTM}}$ is the corresponding peak's d-spacing as reported in the ASTM
card. The stress then is determined by multiplying the strain by the elastic
constant of the material. The calculated strain on the (110) plane of CsBr
changed from -0.0023 to -0.0028 with an elapse of 900~hours. We believe it is 
this stress that contributes to the required energy for bromide's 
disassociation. 

\subsection{Theoretical Modeling}
Metal-insulator core-shell structures and nano-rods 
are known to give 
rise to SPR peaks in absorbance spectra in the visible and near IR region
\cite{isabel, bohren}.
As stated earlier, the SPR peak position and intensity strongly depend on the
metal cluster size, shape, its dielectric constant and that of the surrounding.
Since, the surrounding of the metal clusters in this study is invariant
(CsBr), we 
may use the results here to investigate the contribution of size, shape and
cesium's dielectric constant in SPR's peak position. The systematic variation 
in SPR peaks (fig~1 and fig~2) indicate a systematic variation in one of the
mentioned properties. 

The TEM micrograph (fig~5) shows existence of both cylinderical or nano-rod
structures along with spherical core-shell structures in out sample. Hence, it becomes
important to isolate which of these structures contribute to the SPR peaks in 
the wavelength region of 500-600nm. This can be done using the theoretical 
framework given by Mie \cite{mie} to explain the scattering and absorption 
caused by metal clusters. The Mie theory essential is an application of
Maxwell's equation for electromagnetic plane waves incident on metal
particles. Recent works by Balaji et al \cite{bala} extends Mie
theory to explain scattering and absorption by core-shell structures. We
have extended those calculation schemes on our Cesium Bromide-Cesium
core-shell structures and find theoretically for the grain diamensions that
exist in our samples, there are no or very shallow peaks in the visible
region (fig~8). For the same aspect ratio (${\rm r_{core}/r_{mantel}}$) but
decreasing grain size, we see a blue shift in the shallow peak observed.
This is not in agreement with our experimental observations of fig~1.

As for theoretically calculating the SPR extinction cross-section of
nano-spherical clusters, the Gans theory \cite{gan} is both simple and 
accurate and is given as \cite{link}
\begin{eqnarray}
\sigma_{ext}(\omega)=V\left({2\pi \over
3\lambda}\right)\epsilon_m^{3/2}
\sum_i {\epsilon_2(\omega)1/P_i^2 \over
\epsilon_2(\omega)^2+\left[\epsilon_1(\omega)+\epsilon_m{1-P_i \over
P_i}\right]^2}\label{2}
\end{eqnarray}
where ${\rm \epsilon_1}$ and ${\rm \epsilon_2}$ are the real and imaginary
part of the metal nanocluster's dielectric constant and ${\rm \epsilon_m}$ the 
dielectric constant of the media in which the metal nano-clusters are
embedded in. The depolarising factors \cite{nog} can be easily modified for
cylinderically shaped clusters by taking the aspect ratio (c/a) to be far
less than unity \cite{chanti}. Figure~9 shows the extinction cross-sections calculated for
decreasing average grain size but increasing aspect ratio. For the calculations 
here, we have used the frequency dependent dielectric coefficients
reported by \cite{smith}. The curves and the trend are in agreement with the the
experimental trends shown in figure~1. 

For completeness, we used eqn~(\ref{2}) to see the shift in SPR peak
position with varying ${\rm \epsilon_m}$. This gives the SPR sensitivity to
the medium's refractive index \cite{hchen}. Based on our calculations we find 
the slope between SPR peak position and medium's refractive index to be 
145.1~nm/RIU (fig~10). This small slope would in turn demand Cesium Bromide's 
refractive index to change from 1.66 to values greater than 4.10 to explain the 
results of fig~2(a). Thus, we conclude that the variation of optical properties 
seen with ageing in the Cesium Bromide films can be explained based on 
basis of formation of Cesium nano-rods in it. 

\section{Conclusions}
Experimental data suggests that the stress in the as grown films lead to
formation of defects caused by Bromide atom's displacement. These defects 
start collecting together to form clusters. With the curvature of the
grain's surfaces contributing the maximum stress, the formation of Cesium
at the grain surfaces can be understood. This surface Cesium around Cesium
Bromide not only contributes to the core-shell structures present in the film
but also the a site for formation of Cesium nano-rods. With time
the shell-core grain size decreases along with decrease in the average grain
size of the cylinderical rods. Comparing the results with theoretical
simulations suggests cylinderical grains contribute to SPR peaks in the
visible wavelengths with decreasing grain size accompanied with increasing
aspect ratio (ratio of diameter to length) leading to a red-shift in the
peak position.
 
\section*{Acknowledgment}
The authors would like to express their sincere gratitude to Department of
Science and Technology (DST) India for the financial assitance
(SR/NM.NS-28/2010) given for carrying out this work.

\newpage
\section*{Tables}

{\bf Table 1:} {\sl Compares the presence of various elements (given in weight
percent) in the grain and vesticles in in the electron micrographs. The
chemical composition as measured using EDX attachment with the SEM.}\\
\begin{center}

\begin{tabular}{|| c| c| c||}
\hline 
\hline
  Element & CsBr (at grain) & CsBr (at vesticle)  \\ \hline\hline
  CK  & 18.04  &  49.58  \\ \hline
  OK & 15.72  & 22.29   \\ \hline
   MgK & 02.08  &  01.45   \\ \hline
   SiK & 20.85  &  19.70   \\ \hline
 CaK  & 03.35  &  03.49   \\ \hline
  CsL & 17.66 &  03.50   \\ \hline
BrK  & 22.30 &  00.00  \\ \hline\hline
\end{tabular}
\end{center}

\vskip 1cm
{\bf Table 2:} {\sl Grain size (gs) of CsBr and Cs as calculated from the X-Ray
Diffraction Pattern.}\\
\begin{center}
\begin{tabular}{|| c| c| c||}
\hline 
\hline
  Time (Hrs)  &  gs of CsBr (nm) & gs of Cs (nm) \\ \hline\hline
  22  & 65  &  88 \\ \hline
  112  & 55 &  76 \\ \hline
  794  & 58  &  10 \\ \hline
  1034  & 34  &  -- \\ \hline\hline
\end{tabular}
\end{center}

\newpage

\newpage
\section*{Figure Captions}
\begin{itemize}
\item[Fig 1] The Absorbance spectra of a representative sample (not kept in
dessicator) taken at various intervals. Interestingly, as the SPR's peak 
position shifted to higher wavelengths, the intensity showed presistent 
decrease. The SPR's peak position's shift to higher wavelengths with time 
show a linear trend.

\item[Fig 2] The Absorbance spectra of three different CsBr films varying by
thickness show systematic variation with time. (A) The peak position varies
linearly with time, where the (B) slope (rate at which peak position varies
with time) is directly dependent on film thickness. (C) shows the absorption
intensity also varies with time. 

\item[Fig 3] Field-Emission Scanning Electron Microscope (SEM) micrographs
of (a) sample maintained in dessicator and (b) kept outside. Histograms (c)
and (d) shows the grain size (gs) distribution of sample maintained in dessicator 
and those kept outside, respectively.

\item[Fig 4] Field-Emission Scanning Electron Microscope images show the
sequence of events as CsBr grains break away. Micrograph (a) shows
``vesticles" of Cesium inter-connecting grains which (b) break away with
``vesticle" going with one of the grains. These ``vesticles" fall off (c)
giving cesium rods in the film.

\item[Fig 5] Transmission Electron Microscope (TEM) images confirm nature of 
film throughout the thickness of the film is same as that seen on the surface
using SEM.

\item[Fig 6] Surface Analysis using Electron Diffraction (SAED) of (a) the 
core and (b) the rod.

\item[Fig 7] X-Ray diffraction of CsBr films after aging. Both broad peaks
are deconvoluted to show the Cs (${\rm \bigcirc}$) and CsBr (${\rm
\bigtriangleup}$).

\item[Fig 8] Extinction cross-section of Cesium Bromide-Cesium core-shell
structure calculated (see text) for two different grain sizes, namely (a)
1500 and (b) 1200nm, but same aspect ratio (${\rm r_{core}/r_{mantel}}$). 

\item[Fig 8] Theoretically projected variation of SPR peaks caused by Cesium
nano-rods. Family of curves show a redshift with average grain size
decreasing abid increasing aspect ratio (c/a). The calculations where made using
Gans Model (see text). The simulation follow the same trands of fig~1.

\item[Fig 10] Graph shows the relationship between SPR peak position with
surrounding media's refractive index.

\end{itemize}

\newpage
\section*{Figures}
\begin{figure}[h]
\begin{center}
\epsfig{file=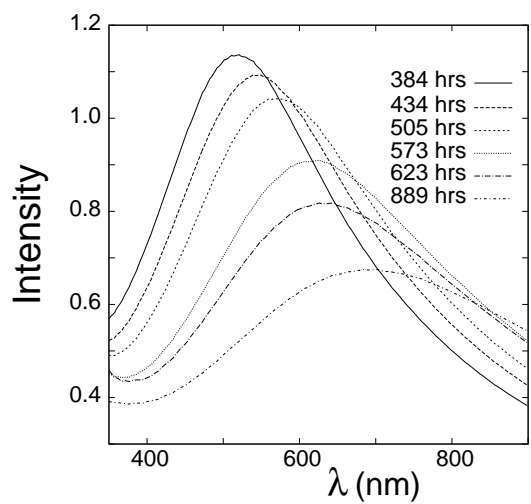,width=2.5in, angle=-90}
\caption{\sl The Absorbance spectra of a representative sample (not kept in
dessicator) taken at various intervals. Interestingly, as the SPR's peak 
position shifted to higher wavelengths, the intensity showed presistent 
decrease. The SPR's peak position's shift to higher wavelengths with time 
show a linear trend.}
\vskip -0.25cm
\end{center}
\vskip -0.25cm
\end{figure}

\newpage

\begin{figure}[h]
\begin{center}
\epsfig{file=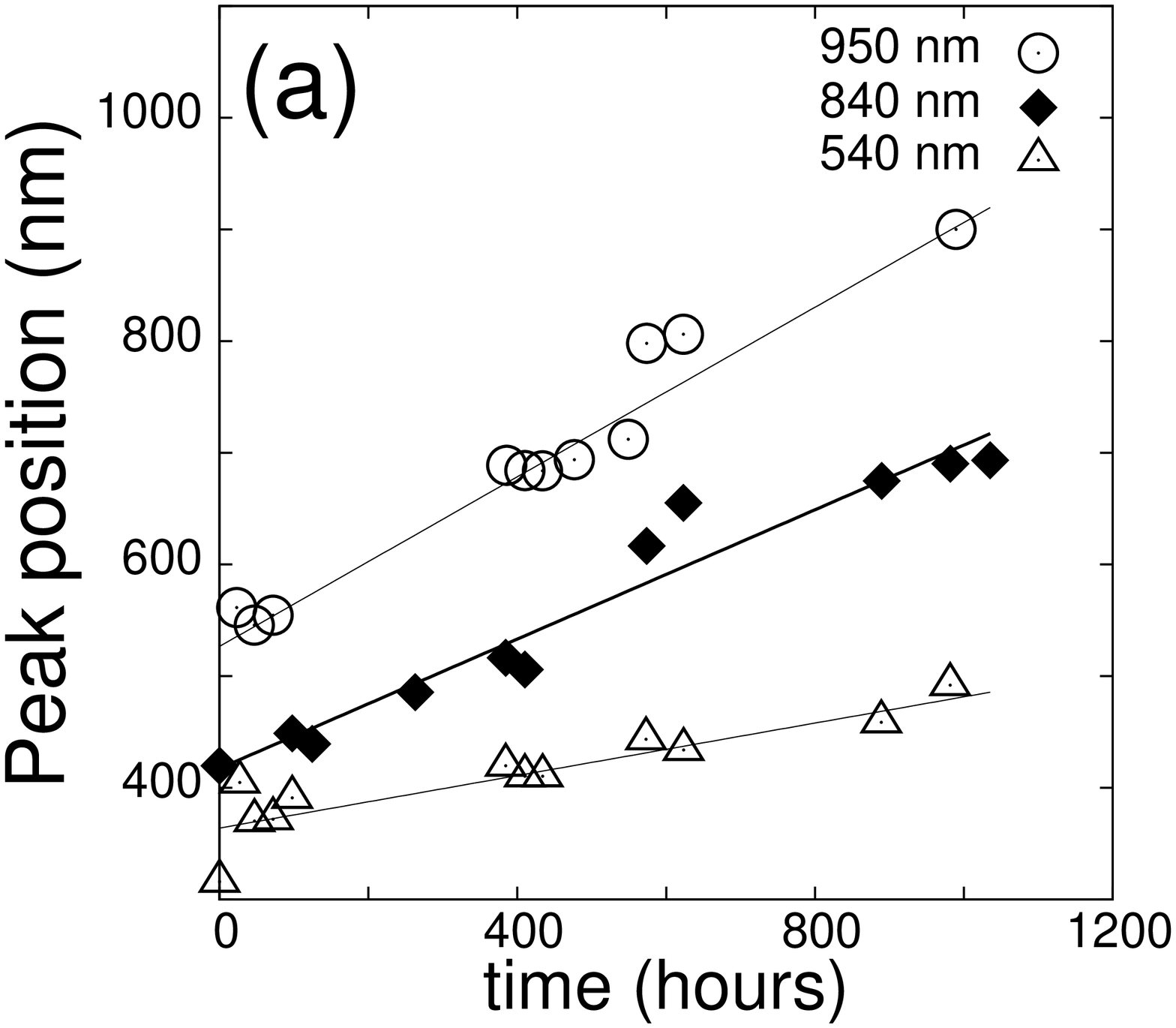,width=2.5in, angle=-90}
\vskip 0.25cm
\vfil
\epsfig{file=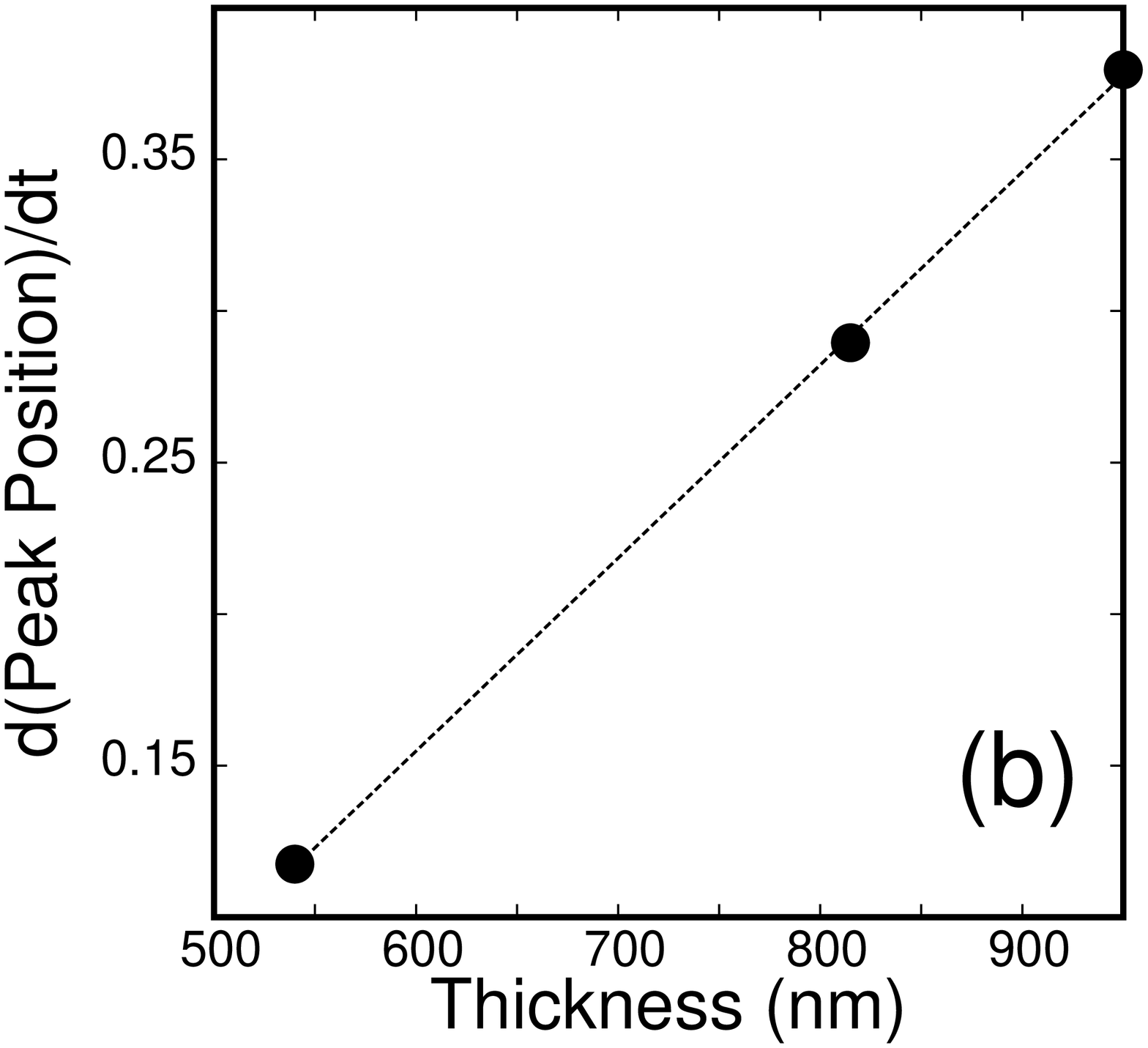,width=2.5in, angle=-90}
\vskip 0.25cm
\vfil
\epsfig{file=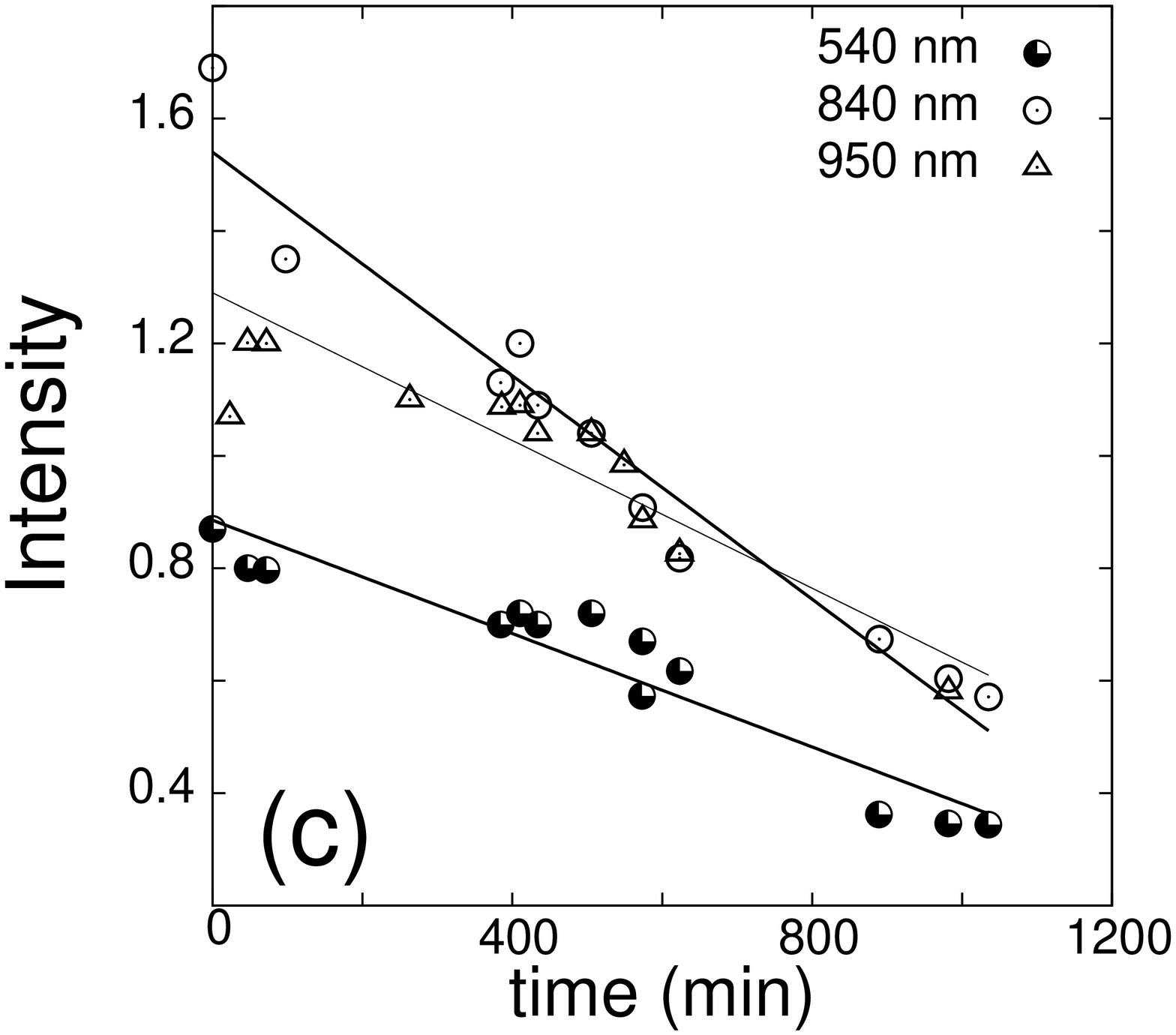,width=2.5in, angle=-90}
\caption{\sl The Absorbance spectra of three different CsBr films varying by
thickness show systematic variation with time. (A) The peak position varies
linearly with time, where the (B) slope (rate at which peak position varies
with time) is directly dependent on film thickness. (C) shows the absorption
intensity also varies with time. }
\vskip -0.25cm
\end{center}
\vskip -0.25cm
\end{figure}
\newpage
\begin{figure}[h]
\begin{center}
\epsfig{file=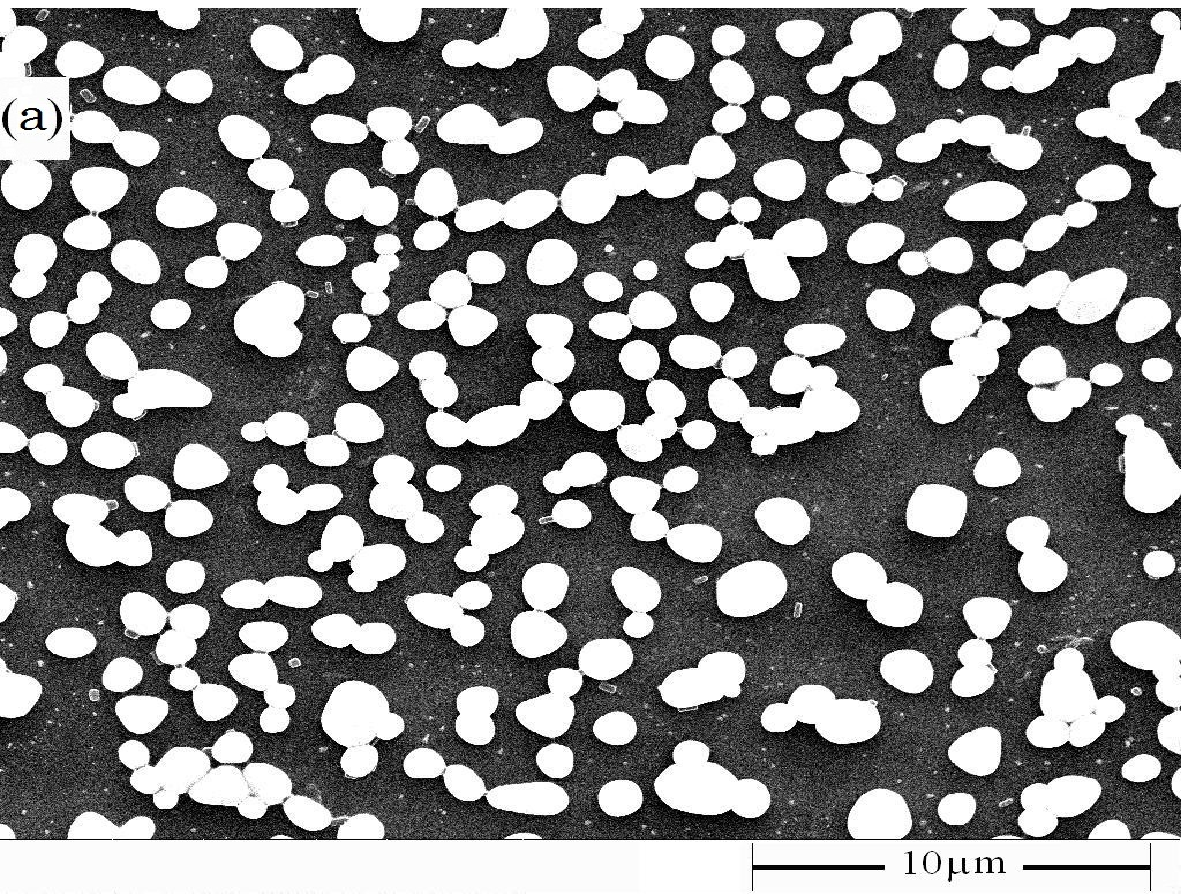,width=2.25in, angle=0}
\hfil
\epsfig{file=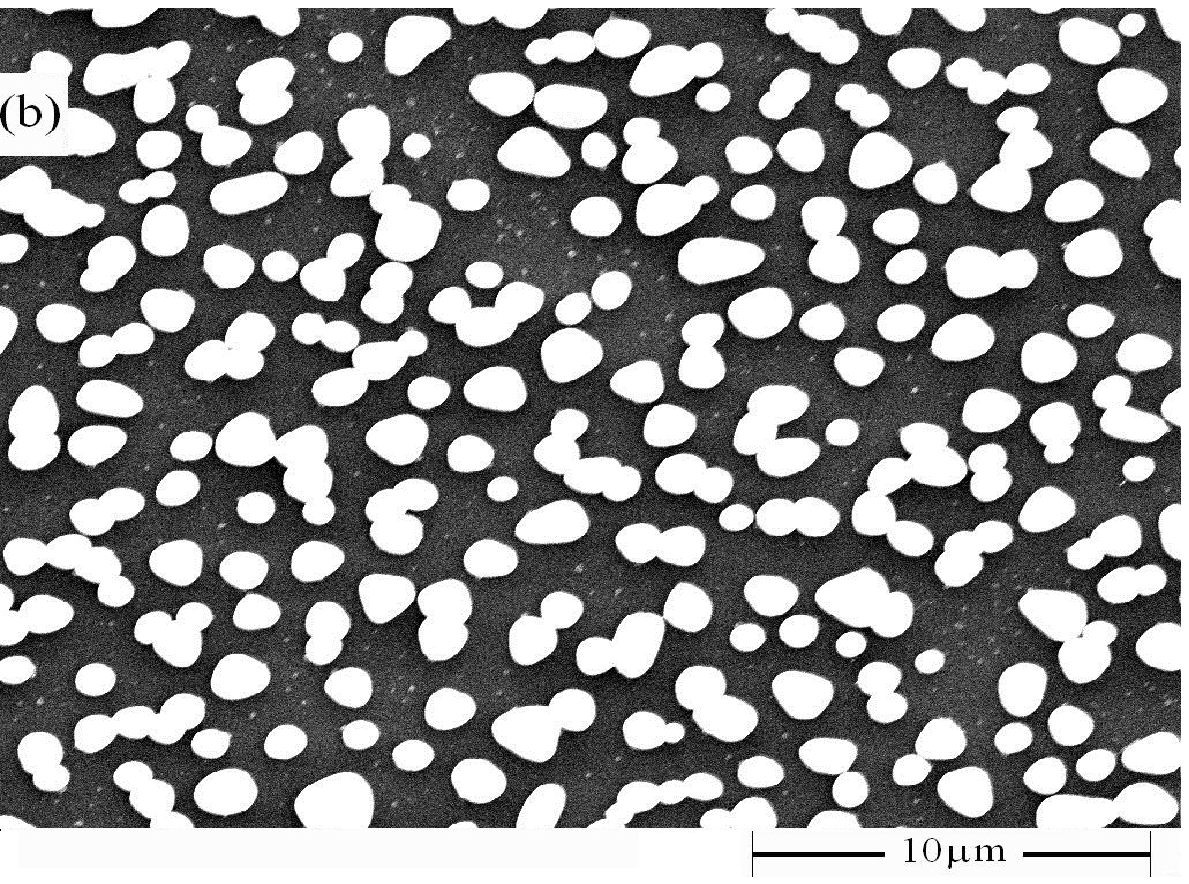,width=2.25in, angle=0}
\vfil
\epsfig{file=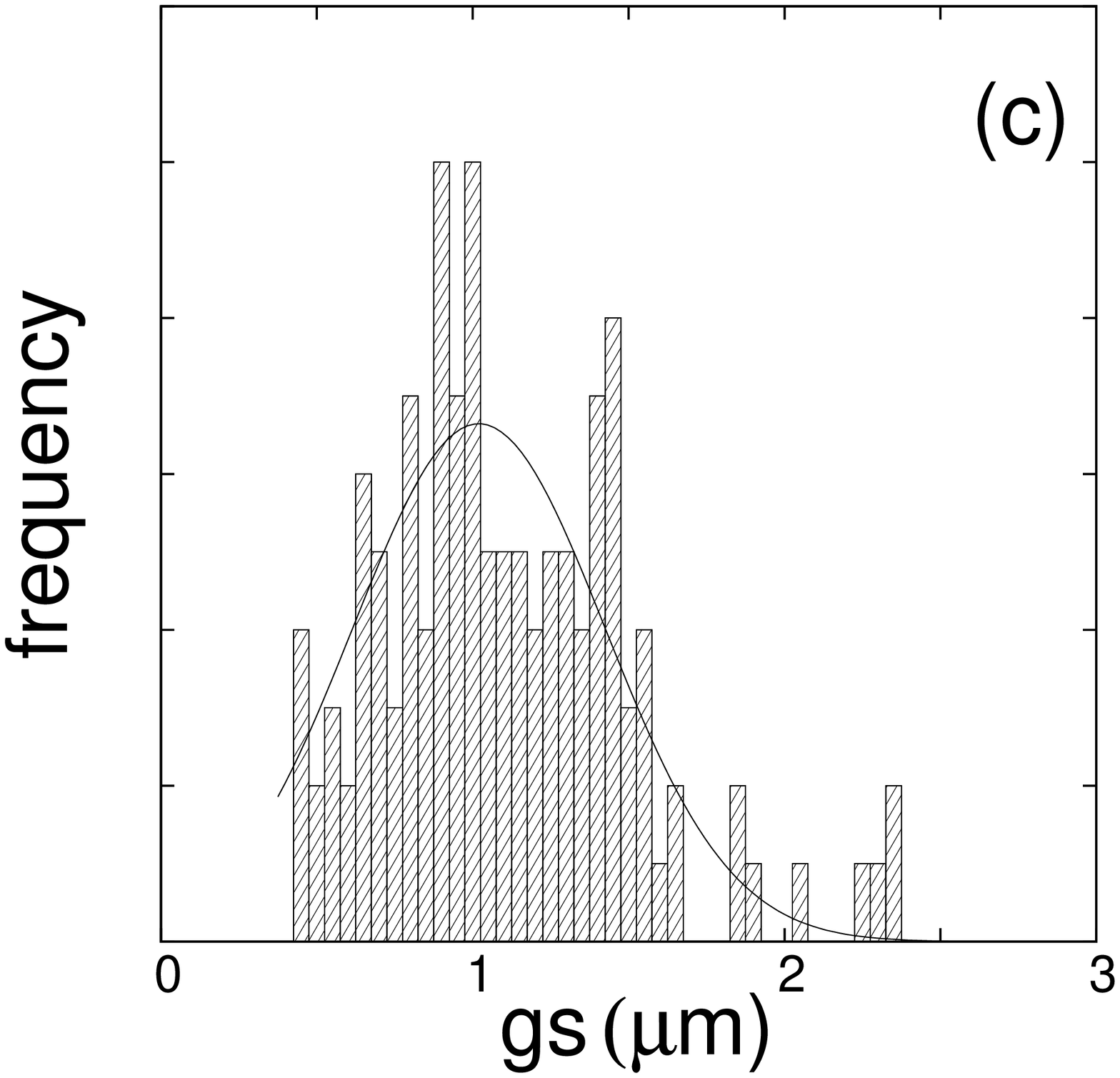,width=2.25in, angle=-90}
\hfil
\epsfig{file=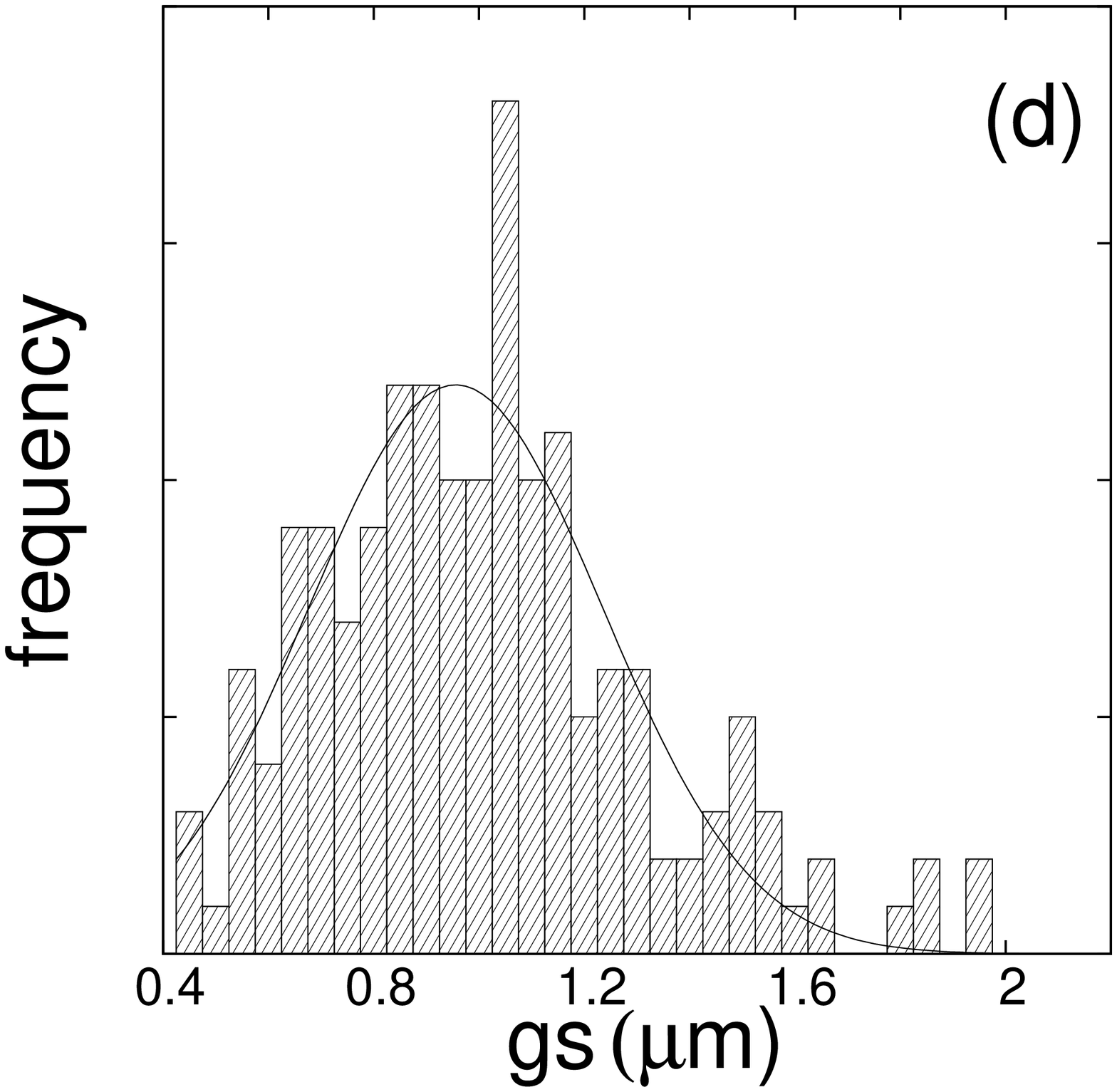,width=2.25in, angle=-90}
\caption{\sl Field-Emission Scanning Electron Microscope (SEM) micrographs
of (a) sample maintained in dessicator and (b) kept outside. Histograms (c)
and (d) shows the grain size (gs) distribution of sample maintained in dessicator 
and those kept outside, respectively.
}
\vskip -0.25cm
\end{center}
\vskip -0.25cm
\end{figure}

\begin{figure}[h]
\begin{center}
\epsfig{file=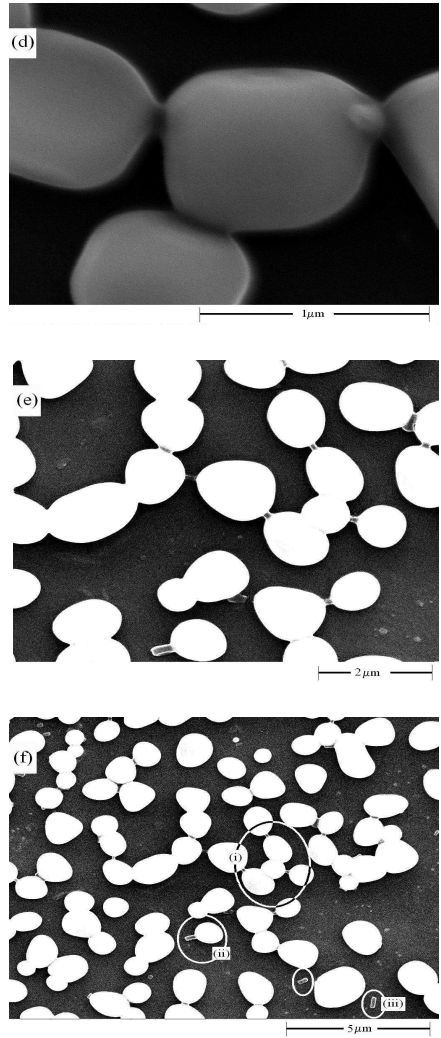,width=2.25in, angle=0}
\caption{\sl Field-Emission Scanning Electron Microscope images show the
sequence of events as CsBr grains break away. Micrograph (a) shows
``vesticles" of Cesium inter-connecting grains which (b) break away with
``vesticle" going with one of the grains. These ``vesticles" fall off (c)
giving cesium rods in the film.
}
\vskip -0.25cm
\end{center}
\vskip -0.25cm
\end{figure}

\begin{figure}[h]
\begin{center}
\epsfig{file=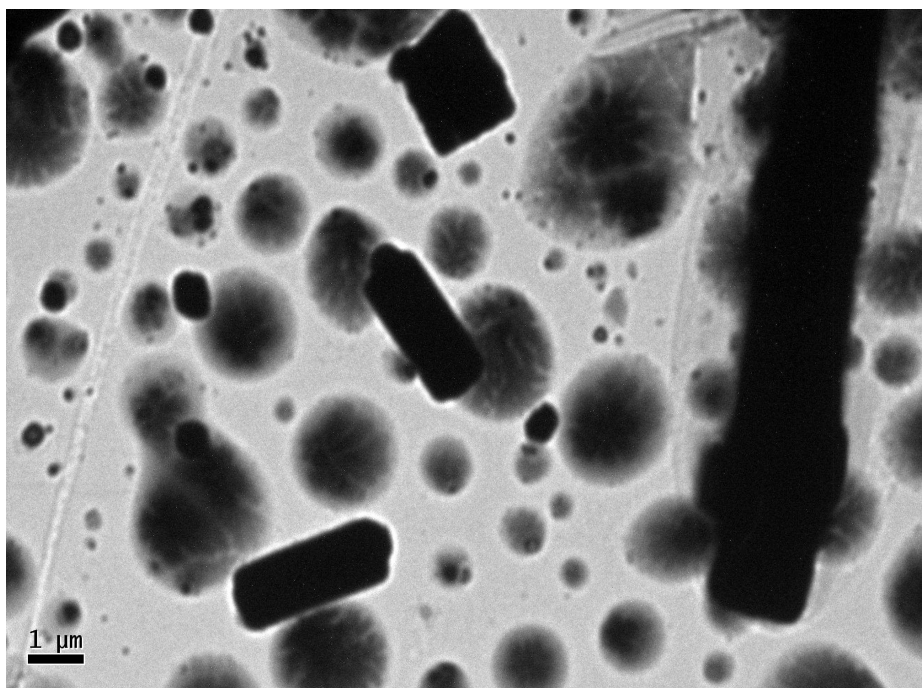,width=4.8in, angle=0}
\caption{\sl Transmission Electron Microscope (TEM) images confirm nature of 
film throughout the thickness of the film is same as that seen on the surface
using SEM.}
\vskip -0.25cm
\end{center}
\vskip -0.25cm
\end{figure}

\newpage

\begin{figure}[h]
\begin{center}
\epsfig{file=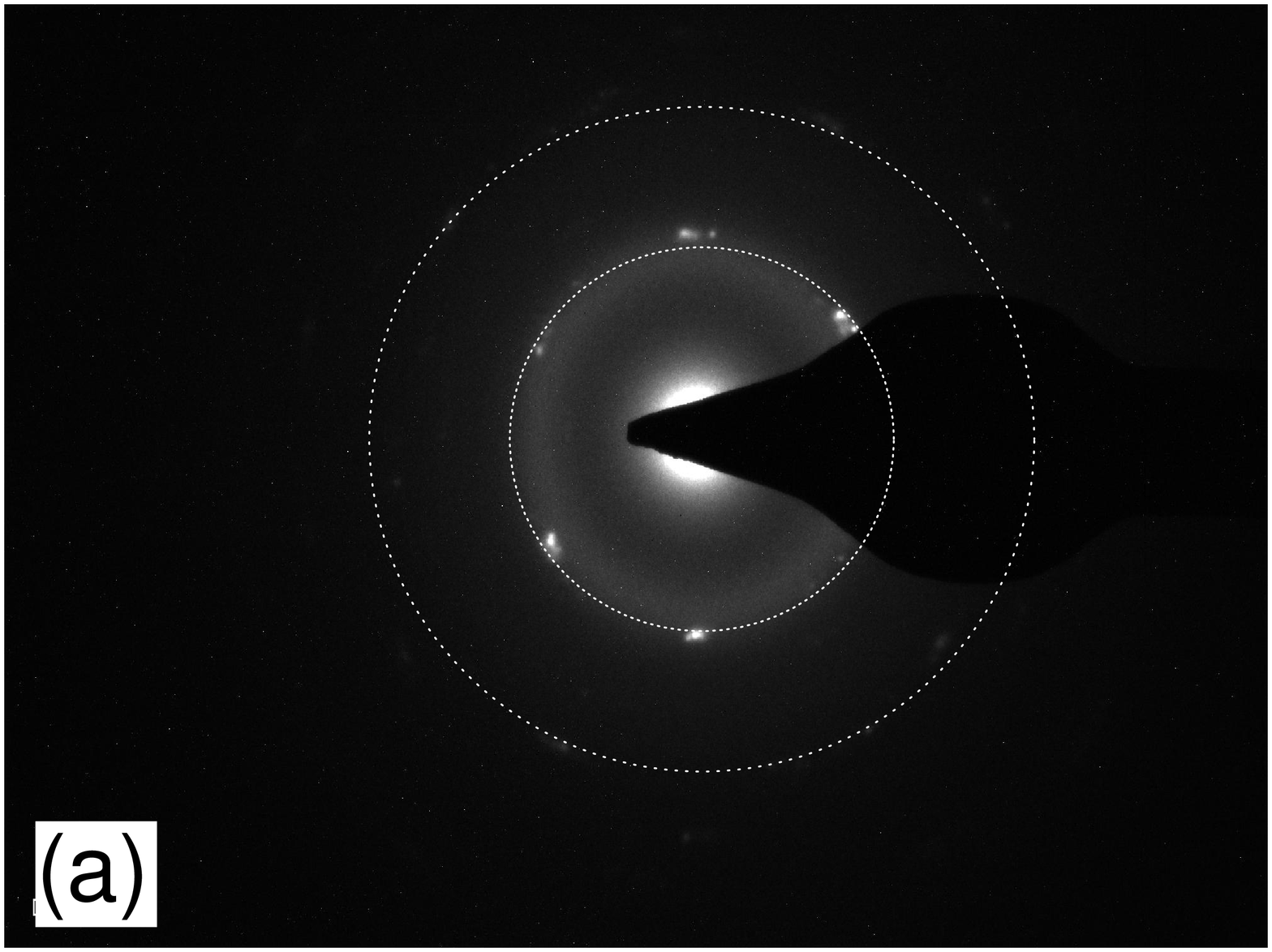,width=4in,angle=0}
\vfil
\vskip 0.3cm
\epsfig{file=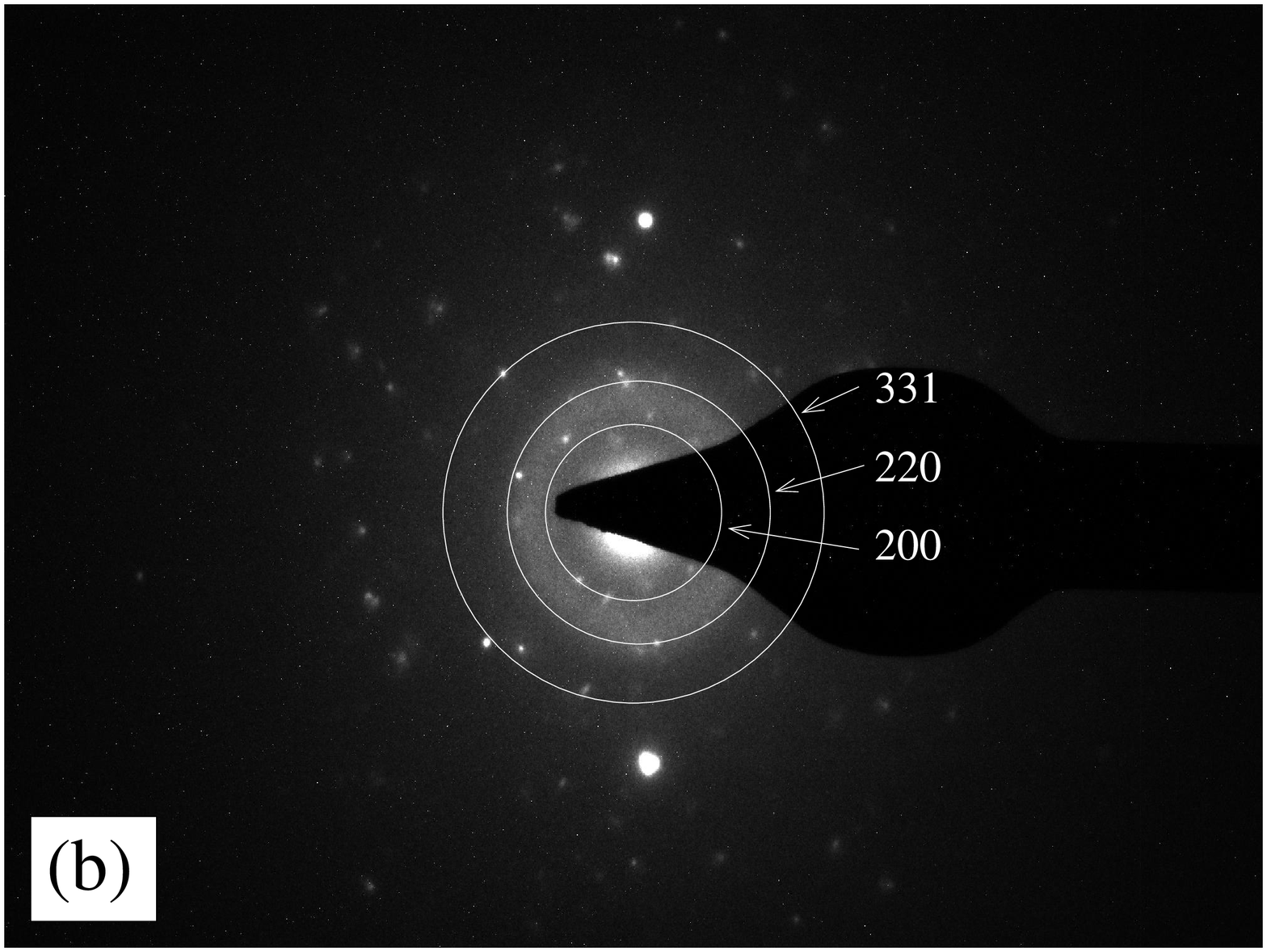,width=4in,angle=0}
\caption{\sl Surface Analysis using Electron Diffraction (SAED) of (a) the 
core and (b) the rod.}
\vskip -0.25cm
\end{center}
\vskip -0.25cm
\end{figure}

\begin{figure}[h]
\begin{center}
\epsfig{file=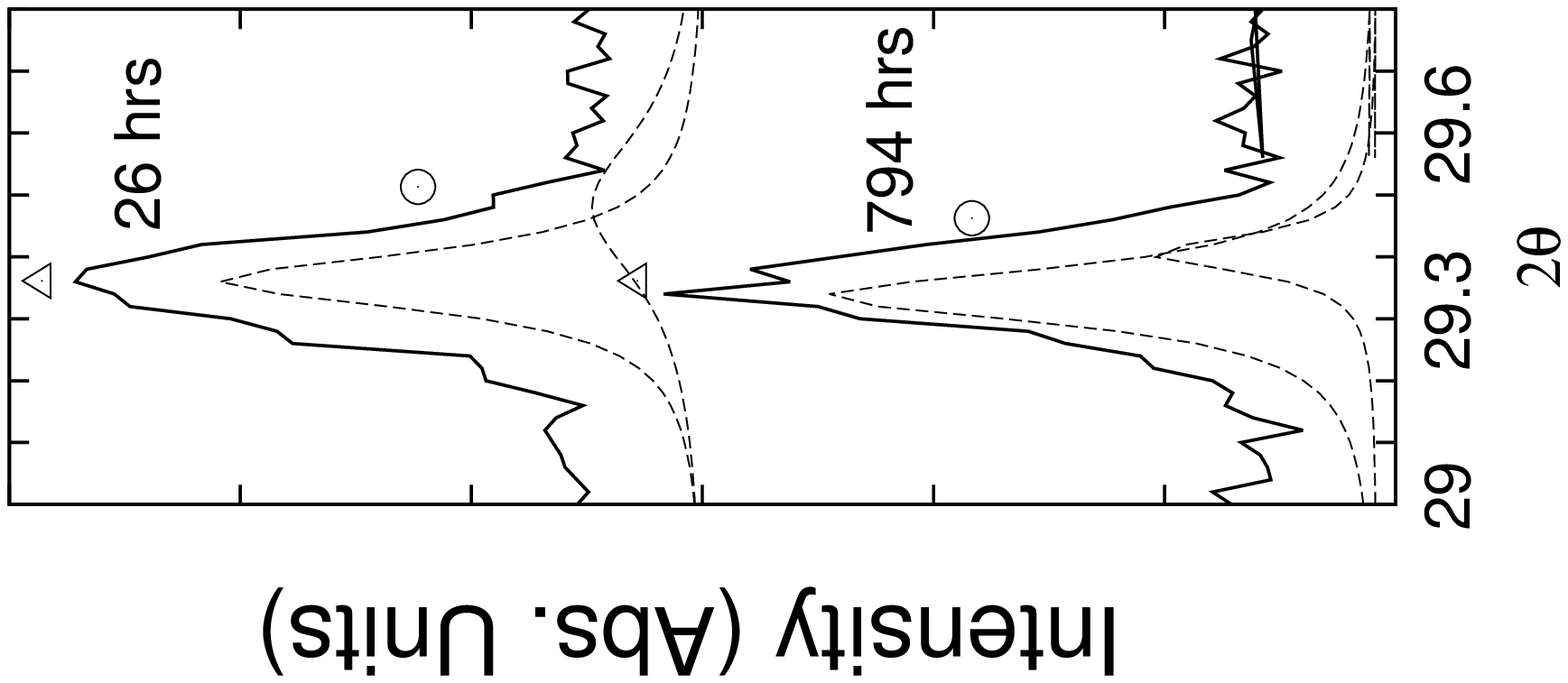,width=4.25in,angle=-90}
\hfil
\epsfig{file=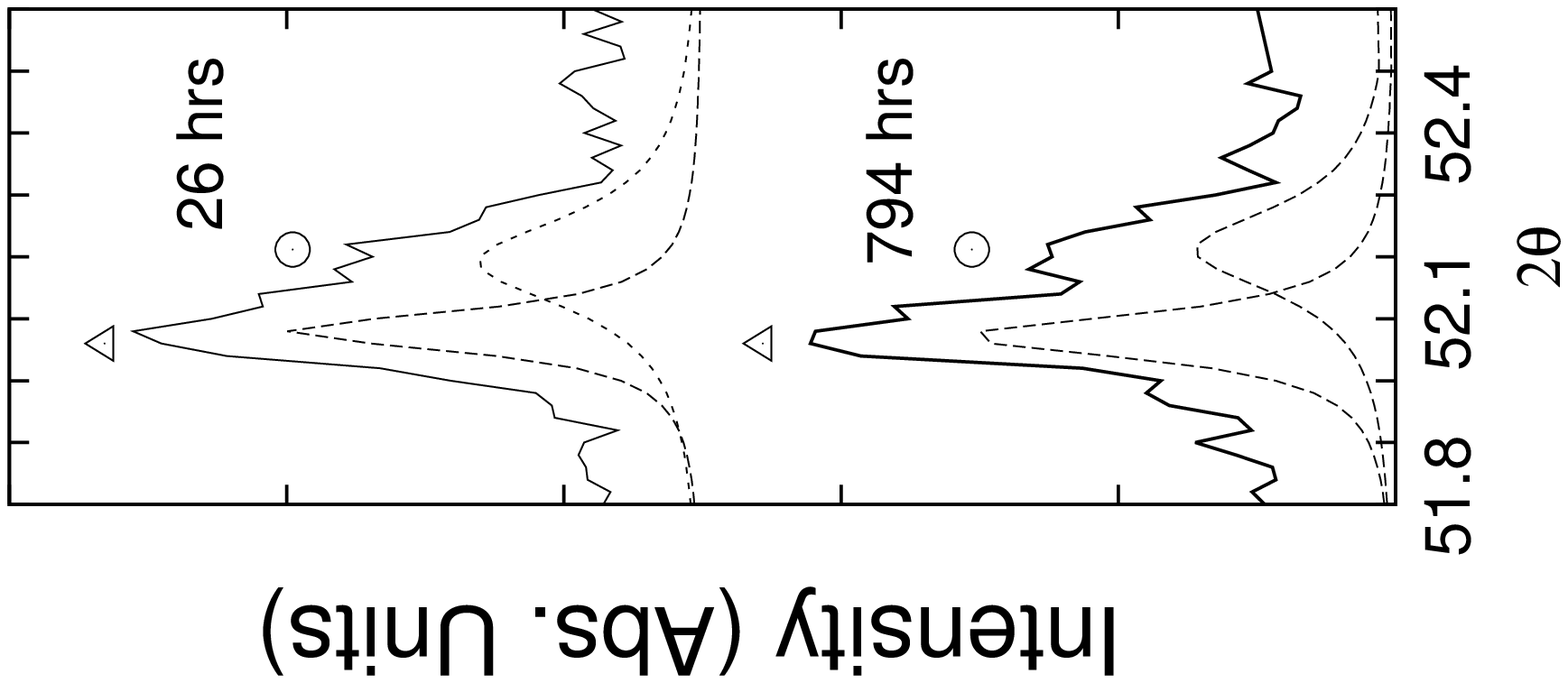,width=4.25in,angle=-90}
\caption{\sl X-Ray diffraction of CsBr films after aging. Both broad peaks
are deconvoluted to show the Cs (${\rm \bigcirc}$) and CsBr (${\rm
\bigtriangleup}$) peaks.}
\vskip -0.25cm
\end{center}
\vskip -0.25cm
\end{figure}

\begin{figure}[h]
\begin{center}
\epsfig{file=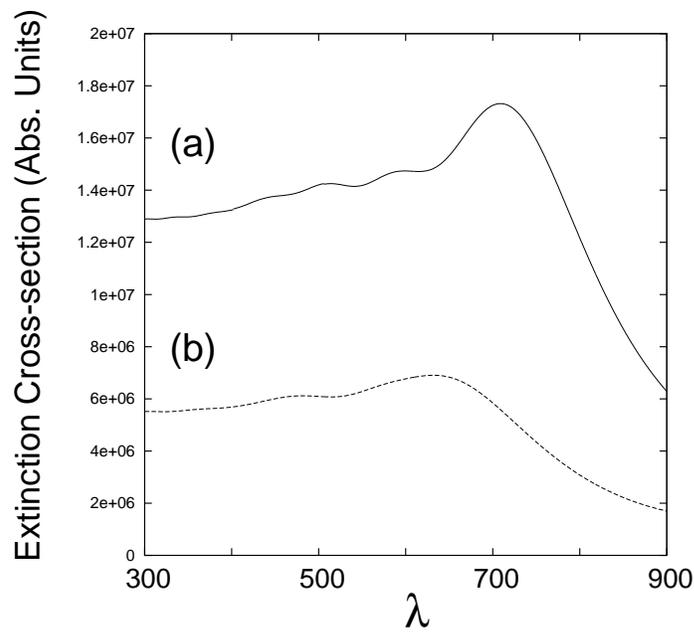,width=3in, angle=-90}
\vskip 0.5cm
\caption{\sl Extinction cross-section of Cesium Bromide-Cesium core-shell
structure calculated (see text) for two different grain sizes, namely (a)
1500 and (b) 1200nm, but same aspect ratio (${\rm r_{core}/r_{mantel}}$). 
}
\vskip -0.25cm
\end{center}
\vskip -0.25cm
\end{figure}

\begin{figure}[h]
\begin{center}
\epsfig{file=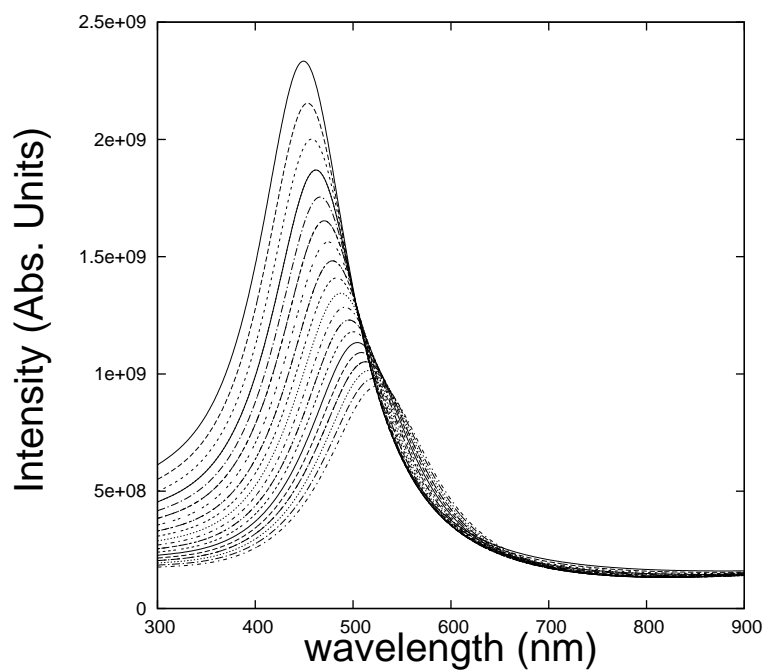,width=3.5in, angle=-90}
\caption{\sl Theoretically projected variation of SPR peaks caused by Cesium
nano-rods. Family of curves show a redshift with average grain size
decreasing abid increasing aspect ratio (c/a). The calculations where made using
Gans Model (see text). The simulation follow the same trands of fig~1.
}
\vskip -0.25cm
\end{center}
\vskip -0.25cm
\end{figure}

\begin{figure}[h]
\begin{center}
\epsfig{file=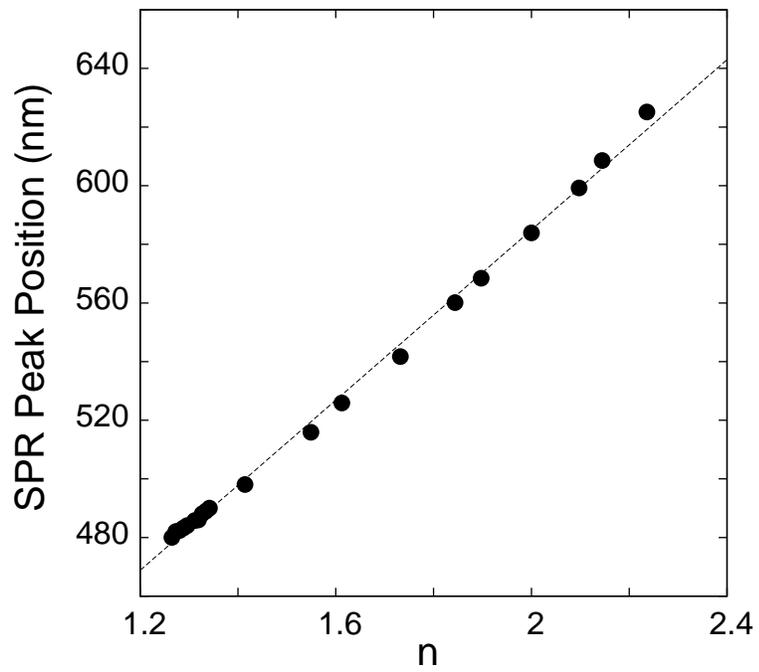,width=3.5in, angle=-90}
\caption{\sl Graph shows the relationship between SPR peak position with
surrounding media's refractive index.
}
\vskip -0.25cm
\end{center}
\vskip -0.25cm
\end{figure}

\end{document}